\documentclass[fleqn,epsf]{myarticle}
\newfont{\mainfont}{cmss12 scaled 1000}      
\newfont{\preprintfont}{cmss12 scaled 1200}      
\newfont{\titlefont}{cmss12 scaled 1500}   
\newfont{\authurfont}{cmcsc10 scaled 1000}   
\newfont{\authuraddfont}{eurm10 scaled 1000} 
\newfont{\Labelfont}{cmss12 scaled 1200}   
\newfont{\labelfont}{cmss12 scaled 1100}   
\newfont{\figfont}{cmss10 scaled 1000}     
\newfont{\footfont}{cmss10 scaled 1000}     
\newfont{\tabfont}{cmss10 scaled 1000}     
\newfont{\reffont}{cmss10 scaled 1000}        
\newfont{\refnamefont}{cmss10 scaled 1000} 
\newfont{\abstractfont}{cmss12 scaled 1000}  
\newfont{\abstractnamefont}{eurm10 scaled 1200} 
\usepackage{amssymb}
\usepackage{epsf}
    \newcommand{\ba}{\begin{eqnarray}}
    \newcommand{\ea}{\end{eqnarray}}
    \newcommand{\be}{\begin{equation}}
    \newcommand{\ee}{\end{equation}}

    \newcommand{\psb}{\bar{\psi}}
    \newcommand{\phd}{\phi^{\dagger}}

\newcommand{\AmS}{{\protect\the\textfont2
  A\kern-.1667em\lower.5ex\hbox{M}\kern-.125emS}}

\title{ \vspace{-2.0cm} \hfill \parbox{40mm} 
{ \preprintfont DESY 96-042 }\\
\vspace{2.0cm}
Implementation of Symanzik's Improvement Program
for Simulations of Dynamical Wilson Fermions in Lattice QCD}

\author{\authurfont{Karl Jansen and Chuan Liu} \\
       \authuraddfont{
                 Deutsches Elektronen 
                 \protect{\rule[0.7mm]{1.2mm}{0.3mm}} Synchrotron DESY} \\
       \authuraddfont{
        Notkestrasse 85, D-22603 Hamburg, Germany }
         }
       
\begin{document}

\maketitle

\begin{abstract}
\setlength{\baselineskip}{0.65cm}
\abstractfont{
We discuss the  
implementation of a Sheikholeslami-Wohlert term
for simulations
of lattice QCD with dynamical Wilson fermions as required
by Symanzik's improvement program.
We show that for the Hybrid Monte Carlo or Kramers equation algorithm
standard even-odd preconditioning can be maintained.
We design tests
of the implementation using analytically and numerically
computed cumulant expansions.
We find that, for situations where the average number of Conjugate Gradient
iterations exceeds 200, the overhead is only
about 20$\%$.}
\end{abstract} 

\newpage

\mainfont       

\section{Introduction}

In Wilson's formulation of lattice QCD, when working at non-vanishing
quark masses, chiral symmetry is violated. Its restoration requires
the approach to the continuum limit and therefore small values of
the lattice spacing $a$. 
Unfortunately, for small values of $a$ and light quark masses 
one needs large lattices and
simulations easily become prohibitively costly.
If, on the other hand, one stays at too large values of the 
lattice spacing, one has to face cut-off effects which are linear
in the lattice spacing in the case of Wilson fermions.
Indeed, in quenched simulations, using Wilson fermions,
the lattice spacing effects were shown to be a severe problem \cite{letter}. 

A well-known remedy of the $O(a)$ effects is anchored in
Symanzik's improvement program \cite{symanzik} which leads to 
adding $O(a)$ correction terms to the lattice action.
Such a term has been worked out for lattice QCD some time ago
\cite{clover} and comes under the name of
a Sheikholeslami-Wohlert (SW) term. 
It has a
coefficient, $c_{sw}$, which 
has to be fixed so as to cancel the $O(a)$ effects.
This kind of improvement has been used already in quenched simulations
(see \cite{quench} for a review).
Recently, using PCAC relations within 
the Schr\"odinger functional setup \cite{letter,sf},  
a non-perturbative determination of the
coefficient
was proposed and, at least in the quenched approximation, 
a significant effect of improvement could be verified.

It is a very natural next step to go beyond the quenched approximation 
and study the effects of
improvement also in dynamical fermion simulations. 
In this paper we want to make a first step in this direction by
describing how the SW-term can be implemented for simulations
of lattice QCD using molecular dynamics algorithms like 
Hybrid Monte Carlo \cite{tony} or Kramers equation \cite{horowitz,kramers}.
An important observation is that for the implementation 
conventional even-odd 
preconditioning
\cite{precond} can be maintained \footnotemark .
\footnotetext{\footfont Of course, also different preconditioning 
techniques like the one proposed in \cite{ssor} could be used.}
We will give analytically and numerically computed cumulant expansions
that can serve as checks of the implementation.
Finally we will discuss 
the crucial
question, whether 
the gain in reducing the lattice spacing effects will merit 
the overhead of adding a SW-term to Wilson's fermion 
action.           

\begin{figure}[t]
\vspace{-0mm}
\centerline{ \epsfysize=7.5cm
             \epsfxsize=10.5cm
             \epsfbox{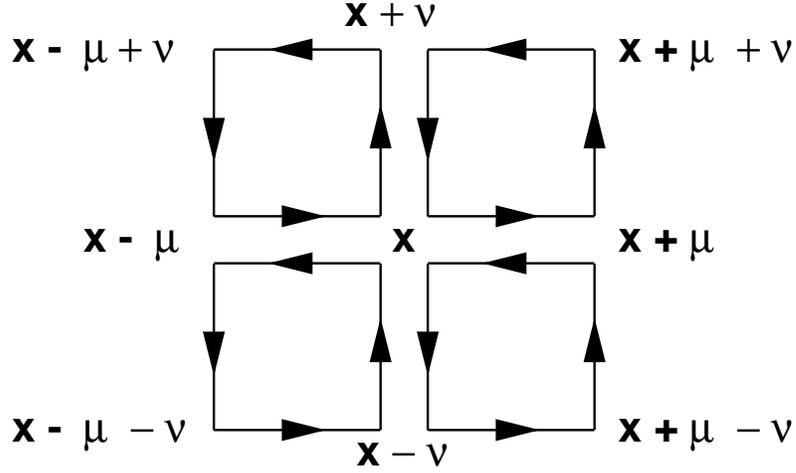}}
\vspace{-0mm}
\begin{center}
\parbox{12.5cm}{\caption{ \label{fig:clover}
The diagram that represents ${\cal F}_{\mu\nu}$.
}}
\end{center}
\end{figure}

The theory is established on a Euclidean space-time lattice with
size $L^3\times T$. With lattice spacing set to unity from now on, the
points on the lattice have integer coordinates $(t,x_1,x_2,x_3)$ which 
are in the range $0\le t < T;0\le x_i < L$. 
A gauge field $U_{\mu}(x)\in SU(3)$ is assigned to the link
pointing from point $x$ to point $(x+\mu)$, where 
$\mu=0,1,2,3$ designates the 4 forward directions in space-time.
The gauge fields assume periodic 
boundary conditions. 
The quark fields $\psi_{Aa\alpha}(x)$ are defined on each
site of the lattice, where $A$,$a$ and $\alpha$ are flavor, 
color and Dirac indices respectively. 
The partition function
for Wilson QCD is given by,
\be
{\cal Z}= \int {\cal D}U{\cal D}\psb{\cal D}\psi
       \exp\;\left(- S_g - S_w -S_{sw} \right)\;\;.
\ee 
The first term $S_g$ is the pure gauge action and is given by
\be
\label{eq:gauge}
S_g = -{\beta \over 6} \sum_{P} Tr(U_P+U^{\dagger}_P)\;\;. 
\ee
The symbol $U_P$ represents the usual plaquette value on the lattice.
The bare gauge coupling $g_0$ is related to $\beta$ by 
$g^2_0 \equiv 6/\beta$.
The second term in the action is the usual Wilson fermion
action:
\be
\label{eq:wilson}
S_w = \sum_{x} \psb(x)(D+m)\psi(x)\;\;,
\ee
where $m$ is the bare quark mass. 
The Wilson difference operator $D$ which appears in the
above expression is given by:
\ba
D&=& {1\over 2}\sum_{\mu} \gamma_{\mu} 
    (\nabla_{\mu}+\nabla^{*}_{\mu})-\nabla_{\mu}\nabla^{*}_{\mu}\;\; , \nonumber\\
\nabla_{\mu}\psi(x)&=& U_{\mu}(x)\psi(x+\mu)-\psi(x)\;\; , \nonumber \\
\nabla^{*}_{\mu}\psi(x)&=& \psi(x)-U^{\dagger}_{\mu}(x-\mu)\psi(x-\mu)\;\;. 
\ea
The third term $S_{sw}$ in the action is the
Sheikholeslami-Wohlert (or clover) term.         
We write it as:
\be
S_{sw} = {i \over 4} c_{sw} \sum_{x,\mu,\nu} 
      \psb(x) \sigma_{\mu\nu} {\cal F}_{\mu\nu}(x) \psi(x)\;\;,
\ee
where $c_{sw}$ is a parameter that has to be tuned such that
this term will eliminate the $O(a)$ effects in the 
on-shell physical observables
 (Symanzik improvement).  To tree level,
this parameter should be set to unity. 
The Dirac matrices $\sigma_{\mu\nu}$ are 
defined in the usual way via $\gamma$-matrices.
The antisymmetric and antihermitian tensor ${\cal F}_{\mu\nu}(x)$ 
is a function of the gauge links and given by:
\newpage
\ba
{\cal F}_{\mu\nu}(x) &=& {1 \over 8} \left[
 U_{\mu}(x)  U_{\nu}(x+\hat{\mu})
U^{\dagger}_{\mu}(x+\hat{\nu})  U^{\dagger}_{\nu}(x)
\right.
\nonumber \\
&+& U_{\nu}(x) U^{\dagger}_{\mu}(x+\hat{\nu}-\hat{\mu})
U^{\dagger}_{\nu}(x-\hat{\mu})  U_{\mu}(x-\hat{\mu})
\nonumber \\
&+& U^{\dagger}_{\mu}(x-\hat{\mu}) U^{\dagger}_{\nu}(x-\hat{\nu}-\hat{\mu})
U_{\mu}(x-\hat{\nu}-\hat{\mu})  U_{\nu}(x-\hat{\nu})
\nonumber \\
&+& U^{\dagger}_{\nu}(x-\hat{\nu}) U_{\mu}(x-\hat{\nu})
U_{\nu}(x-\hat{\nu}+\hat{\mu})  U^{\dagger}_{\mu}(x)
\nonumber \\
&-& \left. h.c. \right]\;\;.
\ea 
We can represent this quantity by its corresponding
diagram (see Fig.~\ref{fig:clover}), 
which suggests the name ``clover''. 
  
We will assume that the mass matrix in equation~(\ref{eq:wilson})
is proportional to the unit matrix in flavor space, namely we
are considering two flavors of Wilson fermions 
with degenerate masses.
The fermion fields in the partition function can be integrated
out, resulting in the fermion determinant. This determinant is then
``integrated in'', using the pseudofermion (boson)
 fields $\phi^{\dagger}(x)$
and $\phi(x)$ for the convenience of
the simulation. 
We then write the partition function and effective action of the theory as:
\ba
\label{eq:mdm}
{\cal Z}&=&\int {\cal D}U{\cal D}\phd{\cal D}\phi
       e^{- S_{eff}}\;\; ,\nonumber \\
S_{eff}&=&S_g + \phi^{\dagger}Q^{-2} \phi \;\;,
\ea
where the matrix $Q$ is related to the fermion matrix $M$ by 
$Q=c_0\gamma_5M$ and, including the SW-term, 
is given by:
\ba
\label{eq:seff}
Q(U)_{xy} \!\!\!&=& \!\!\!c_0\gamma_5 [ 
(1+{i \over 2}c_{sw}\kappa\sigma_{\mu\nu}{\cal F}_{\mu\nu}(x))\delta_{x,y}
 \nonumber \\
&-&\kappa\sum_{\mu} \{
   (1-\gamma_{\mu})U_{\mu}(x)\delta_{x+\mu,y} + 
(1+\gamma_{\mu})U^{\dagger}_{\mu}(x-\mu)\delta_{x-\mu,y}\}]  \;\;,
\ea
with $\kappa=(8+2m)^{-1}$ and $c_0=(1+8\kappa)^{-1}$.
The notation $\sigma_{\mu\nu}{\cal F}_{\mu\nu}$ in the
above expression, and henceforth, implies the summation over $\mu$ and $\nu$.

\section{Implementation of the SW-term}
\label{sec:hmc}
 
In this section we will discuss how the SW-term is implemented for
simulations using Hybrid Monte Carlo or Kramers equation algorithms.
Although
these two algorithms show a comparable performance,
the Kramers equation algorithm seems to be
advantageous due to possible problems of non-reversibility
in the Hybrid Monte Carlo algorithm \cite{kramers}. 
Since the
use of preconditioning techniques in
fermion simulations appears to be very important
we will discuss directly how 
to install the SW-term with, in our case, even-odd preconditioning, see also
\cite{luo} for a similar discussion.
Let us write the matrix $Q$ in equation~(\ref{eq:seff}) as
\be
Q \equiv c_0 \gamma_5  \left( \begin{array}{cc}
                1+T_{ee} & M_{eo} \\
                M_{oe} & 1+T_{oo} \\
                \end{array} \right)\;\;, 
\ee
where we have introduced the matrix $T_{ee}$($T_{oo}$) on the
even (odd) sites as
\be \label{eq:t}
(T)_{xa\alpha,yb\beta} = 
{i \over 2}c_{sw}\kappa\sigma^{\alpha\beta}_{\mu\nu}
{\cal F}^{ab}_{\mu\nu}(x) \delta_{xy}\;\;.
\ee
The off-diagonal parts $M_{eo}$ and $M_{oe}$, which 
connect the even with odd and odd with even lattice 
sites respectively, are just
the conventional Wilson hopping matrices.
Preconditioning is now realized by writing the determinant of $Q$,
apart from an irrelevant constant factor, as
\ba
det(Q)&\propto&det(1+T_{ee})det\hat{Q}
\nonumber \\
\hat{Q}&=&\hat{c}_0 \gamma_5(1 + T_{oo} - M_{oe}(1+T_{ee})^{-1}M_{eo})\;\;.
\ea
The constant factor 
$\hat{c}_0$ is given by $\hat{c}_0=1/(1+64\kappa^2)$. 
For the simulation with even-odd preconditioning, 
 one introduces the Hamiltonian: 
\be
\label{eq:ham}
{\cal H} = \sum_{x,\mu}{1 \over 2}Tr(H^2_{\mu}(x)) 
+ S^{(eo)}_{eff}(U_{\mu},\phd,\phi)\;\;,
\ee
where $H_{\mu}(x)$ represents the momentum conjugate to the 
gauge field $U_{\mu}(x)$ and 
takes values in the Lie algebra of $SU(3)$, i.e. a 
traceless hermitian $3\times 3$ matrix. 
The effective action $S^{(eo)}_{eff}$ for the even-odd
preconditioned case is given by:
\ba \label{eoaction}
S^{(eo)}_{eff} &=& S_g[U_{\mu}]+ S_{det}[U_{\mu}]+S_{b}[U_{\mu},\phd,\phi]\;\;,
\nonumber \\
S_g[U_{\mu}] &=& -(\beta/6)\sum_{P}Tr(U_{P} +U^{\dagger}_{P})\;\;,
\nonumber \\
S_{det}[U_{\mu}] &=& -2Tr\log(1+T_{ee}) \;\;,
\nonumber \\
S_{b}[U_{\mu},\phd,\phi] &=& \phd \hat{Q}^{-2} \phi\;\;.
\ea
The first term $S_g$ in $S_{eff}^{(eo)}$ is just the usual Wilson pure gauge
plaquette action. The second term involves the determinant
of the matrix $(1+T_{ee})$ on the even lattice sites.
We shall refer to it as the determinant contribution.
The third term is the determinant of the preconditioned 
matrix $\hat{Q}$ which acts only on the boson fields at odd lattice sites.
We will call $S_b$
the bosonic contribution.

The preconditioned matrix $\hat{Q}$ contains the inverse matrix
$(1+T_{ee})^{-1}$ which is needed only locally at a given lattice point $x$.
Separating the lower and upper components of the fermion fields, 
the computation of the inverse amounts to the inversion of two 
complex $6\times 6$
matrices per lattice site. 
This can be done using, for example,
 a Householder triangularization \cite{house}
which is advantageous on parallel computers, especially on the Alenia
Quadrics (APE) machines we are using here \cite{ulli}.
In order to study, whether there might occur problems with inverting
the matrix $(1+T_{ee})$, we measured the lowest value $D_{min}$ of 
the determinants  
$det(1+T_{ee}(x))$ for every configuration. As an example, 
for $\beta=5$, $\kappa = 0.150$ and
$c_{sw} =1.52$ we find that $D_{min}=0.584(2)$. 
Noting that the critical hopping parameter $\kappa_c$ is lowered
for SW-improved fermions as compared to conventional
Wilson fermions \cite{allton} and,  
since for larger
values of $\beta$ and(or) lower values of $\kappa$, $D_{min}$
is increasing, we do not expect problems with the matrix inversion
for realistic QCD simulations.

Since the bosonic part $S_b$ is quadratic 
in the $\phi$ fields, they are generated at the
beginning of each molecular dynamics trajectory via
\be
\phi= \hat{Q}R\;\;,
\ee
where $R$ is a random spinor field taken from a Gaussian distribution
of norm one. 
The gauge fields and its
corresponding momenta are updated according to:
\ba
\label{eq:update}
 U_{\mu}^{'}(x) &=&\exp(iH_{\mu}\delta t)U_{\mu}(x)\;\;, \nonumber \\
i \delta H_{\mu}(x)&=& [U_{\mu}(x)F_{\mu}(x)]_{T.A.}\delta t \;\;,
\ea
where $[\cdots]_{T.A.}$ stands for the traceless
antihermitian part of the matrix \cite{sugar} and
the quantity $[U_{\mu}(x)F_{\mu}(x)]_{T.A.}$ is the ``total force''
associated with the link $U_{\mu}(x)$ which will
be discussed in more detail below.
In the simulation the update eq.(\ref{eq:update}) is realized via
a leapfrog integration scheme.

In molecular dynamics algorithms, $F_{\mu}$ is the
``coefficient'' in the change of the
effective action when an infinitesimal change of the
gauge link $\delta U_{\mu}(x)$ is applied, i.e.
\be
\label{eq:linear}
\delta S^{(eo)}_{eff} = \sum_{x \mu} 
Tr(F_{\mu}(x)\delta U_{\mu}(x) +F^{\dagger}_{\mu}(x)\delta U^{\dagger}_{\mu}(x))\;\;.
\ee
The quantity $F_{\mu}(x)$ receives three contributions. 
The first one, $V_{\mu}(x)$, is coming
from the pure gauge part. The second
is from the determinant of the SW-term on the even sites.
The third one is coming from the change in $\phd\hat{Q}^{-2}\phi$. 
We will now discuss the last two terms in some detail.

Let us start with the third contribution, namely the variation of
the effective action from the bosonic term $S_b$.
A direct variation of this term gives:
\be
\delta S_b = -(X^{\dagger}_o \delta \hat{Q} Y_o
             +Y^{\dagger}_o \delta \hat{Q} X_o)\;\;,
\ee
where $X_o \equiv \hat{Q}^{-2} \phi_o$ 
and $Y_o \equiv \hat{Q}^{-1} \phi_o$
are fields defined only on odd lattice sites.
They are obtained by inverting the preconditioned
matrix $\hat{Q}^2$ with a given source $\phi_o$.

Using          
\ba
\delta \hat{Q}&=&\hat{c}_0 \gamma_5 \left[ 
  \delta T_{oo} - \delta M_{oe}(1+ T_{ee})^{-1}M_{eo}
   -  M_{oe}(1+ T_{ee})^{-1}\delta M_{eo} \right.
\nonumber \\
    &+& \left. M_{oe}(1+ T_{ee})^{-1} \delta T_{ee}(1+T_{ee})^{-1}M_{eo} \right] \;\;,
\ea
we find that the variation $\delta S_b$ can be written as
\be
\delta S_b = - (\hat{c}_0/c_0)(X^{\dagger} \delta Q Y
               +Y^{\dagger} \delta Q X ) \;\;,
\ee
where now, the fields $X$ and $Y$ are defined over the full
lattice. 
For the field $X$, we have:
\be
X = \left( \begin{array}{c}
           -(1+T_{ee})^{-1} M_{eo} X_o \\
           X_o \\
           \end{array} \right) \;\; ,
\ee
and similarly for the field $Y$. 
The variation of the original matrix $Q$ again consists of two parts.
The contribution from the off-diagonal elements is the same as
in the conventional Wilson case. 
This term will be denoted as 
$F^{(w)}_{\mu}(x)$ for a given link $U_{\mu}(x)$.
Therefore, the only new ingredient is from the diagonal terms
which explicitly involves the contributions of the SW-term.
To obtain this contribution, let us focus on the variation of one particular
link variable $U_{\mu}(x)$, keeping
all others constant, and try to collect all possible 
contributions.  One easily finds out that, in order for
the ``clover leaves'' of the SW-term to contain  the chosen link 
$U_{\mu}(x)$, the center of the clover can only be 
at lattice points $x$, $x+\mu$, $x \pm \nu$ and $x+\mu \pm \nu$,
where $\nu$ is some direction different from $\mu$.
\begin{figure}[t]
\vspace{-0mm}
\centerline{ \epsfysize=7.5cm
             \epsfxsize=12.5cm
             \epsfbox{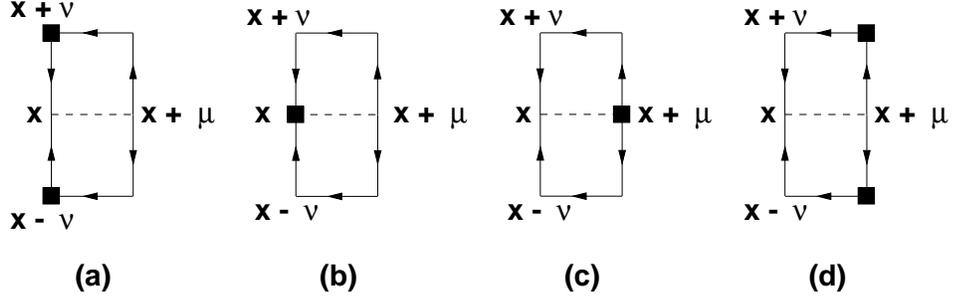}}
\vspace{-0mm}
\begin{center}
\parbox{12.5cm}{\caption{ \label{fig:diagram}
The diagrams that contribute to $F_{\mu}^{(sw)}(x)$, eq.(\ref{eq:f_sw}).
}}
\end{center}
\end{figure}
After realizing this, it is just a matter of book-keeping to gather
all the possible terms. We will represent them collectively
in a diagrammatic fashion, see Fig.~\ref{fig:diagram}. 
We write the contribution, $F^{(sw)}_{\mu}(x)$, from these diagrams as:
\be
\label{eq:f_sw}
F^{(sw)}_{\mu}(x) = - {\hat{c}_0c_{sw}\kappa \over 8}
 ({\mbox{ \mainfont diagrams in Fig.~2 }})\;\;.
\ee
The diagrammatic representations have the following
simple interpretation. Starting from point $x+\mu$,
one can, following the direction of 
the arrows on the lines, go around the plaquette
and finally arrive at point $x$. 
In doing this, one performs matrix multiplications
along the way.  Each line segment with an arrow on
it represents the corresponding gauge link at that point.
A black square at a given point, with
a particular choice of $\mu$ and $\nu$, stands for
an ``insertion'' of a $3\times 3$ matrix in color space.
This matrix is  given by the following expression:
\be
\label{eq:block}
(\blacksquare)_x= Tr_{Dirac}( i\gamma_5 \sigma_{\mu\nu}Y(x)\otimes X^{\dagger}(x)
                   +i\gamma_5 \sigma_{\mu\nu}X(x)\otimes Y^{\dagger}(x) )\;\;.
\ee
The trace in the above formula, as indicated, 
is done only in Dirac space and
$X\otimes Y^{\dagger}$ stands for a direct product in color 
space. 
For a fixed $\mu\nu$ pair, one can take two different
paths to arrive at point $x$, namely the ``upward'' path, which
passes through point $(x+\nu)$, and the ``downward'' path, which
passes through point $(x-\nu)$.
The path that winds downward 
gets a relative minus sign for
its contribution. Finally, one  
should sum up all the contributions from
all possible $\nu \neq \mu$ values.

Let us turn now to the determinant contribution from the even sites.
We have, when taking the variation of the action,
\be
\delta S_{det} = -2 Tr(1 + T_{ee})^{-1} \delta T_{ee}\;\;.
\ee
This contribution can also be easily represented by a set
of diagrams. In fact, by almost the same set of diagrams as
\begin{figure}[t]
\vspace{-0mm}
\centerline{ \epsfysize=7.5cm
             \epsfxsize=12.5cm
             \epsfbox{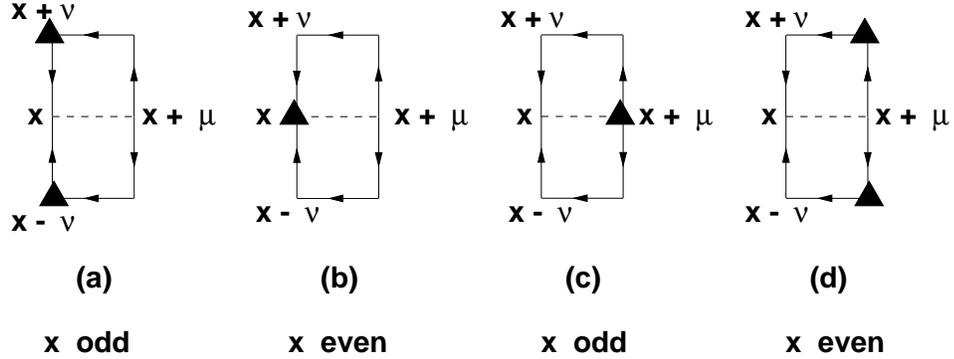}}
\vspace{-0mm}
\begin{center}
\parbox{12.5cm}{\caption{ \label{fig:triangle}
The diagrams that contribute to $F_{\mu}^{(det)}(x)$, eq.(\ref{eq:f_det}).
They are only included when
the insertion points are even lattice points, as indicated
in the label.
}}
\end{center}
\end{figure}
in the previous case, except for the following changes.
First the black square insertions at various points are
replaced by the black triangles at the same points.
And each black triangle, at some chosen point
$x$ and direction $\nu \neq \mu$, represents the following expression:
\be
\label{eq:triangle}
(\blacktriangle)_x= Tr_{Dirac}[ i \sigma_{\mu\nu}(1+T_{ee}(x))^{-1}]\;\;.
\ee
The second modification is that these insertions  should only
be done at even lattice points, otherwise the contribution
is set to zero.
With these modifications, the contribution from
the determinant $F^{(det)}_{\mu}(x)$ 
is given by
\be \label{eq:f_det} 
F^{(det)}_{\mu}(x) 
= -{\kappa c_{sw} \over 4}
 ({\mbox{ \mainfont diagrams in Fig.~3 }})\;\;.
\ee
Finally, $F_{\mu}(x)$ is written as a sum of all the 
contributions as:
\be
F_{\mu}(x)=V_{\mu}(x)+F^{(w)}_{\mu}(x) + F^{(det)}_{\mu}(x) + F^{(sw)}_{\mu}(x) \;\;.
\ee

\mainfont
Let us mention, that 
even-odd preconditioning
can also be achieved in a manner slightly different from the case
just discussed. In this case one treats the
even and odd points more symmetrically. This amounts to 
rewriting the determinant of the fermion matrix as:
\ba
det(M)&=&det(1+T_{ee})det(1+T_{oo})det\hat{M}
\nonumber \\
\hat{M}&=&(1-(1+T_{oo})^{-1}M_{oe}(1+T_{ee})^{-1}M_{eo})\;\;.
\ea
Then, the determinant of $\hat{M}^{\dagger}\hat{M}$
is realized by introducing the bosonic Gaussian fields
while the determinants of $(1+T_{ee})$ and $(1+T_{ee})$  
could be treated the same way as discussed before. 
In this version
the determinant
contributions are then summed up on all lattice
sites, even and odd.
At present, it is not clear, whether one of the two possibilities
of preconditioning the fermion matrix should be preferred.

\section{Tests of the implementation}

Applying an infinitesimal change to the gauge links, the
subroutine which evaluates $F_{\mu}(x)$ was first checked using 
eq.(\ref{eq:linear}) by computing the change of the action explicitly
and by using $F_{\mu}(x)$. 
Besides this, 
the focus of our additional tests will concern 
the SW-term alone.
For this purpose, it is
advantageous to study cases
in which the effects of the SW-term have been singled out.
The SW-term that enters the simulation program only
depends on the parameter combination $\bar{c}_{sw} \equiv c_{sw}\kappa$.
Therefore, one can take the limit $\kappa \rightarrow 0$ but
keep $\bar{c}_{sw}$ at some prescribed value. In this way,
the effects of the Wilson hopping matrices are completely
switched off and the fermionic part of the action depends only on the
SW-term. Of course, the pure gauge parameter $\beta$ is
still at our disposal and the SW-term can be tested at 
various values of $\beta$.

\subsection{Tests at $\beta =0$}

As a first test, we study the model in the limit of
$\beta=0$ and $\kappa=0$, however, 
$c_{sw}\kappa$ is kept finite.
In this limit, the effective action of the theory is simply:
\be
\label{eq:seffb0}
S_{eff} = -2 \sum_{x} Tr\log(1 + T(x)) \;\;,
\ee
with $T(x)$ as given in eq.(\ref{eq:t}).
The advantage of taking this limit is that, 
in this simplified case, the average plaquette value can be
evaluated analytically as a power series expansion in terms
of $\bar{c}_{sw}$.  
For the expectation value of a plaquette, 
$P\equiv (1/6)tr(U_P+U^{\dagger}_P)$, 
to lowest order one gets
\be
\label{eq:p2}
<P> = - {1 \over 12}\bar{c}_{sw}^2 + O(\bar{c}_{sw}^4) \;\;.
\ee
To work out the next leading correction of this expansion requires
\begin{figure}[htp]
\vspace{-0mm}
\centerline{ \epsfysize=13.5cm
             \epsfxsize=13.5cm
             \epsfbox{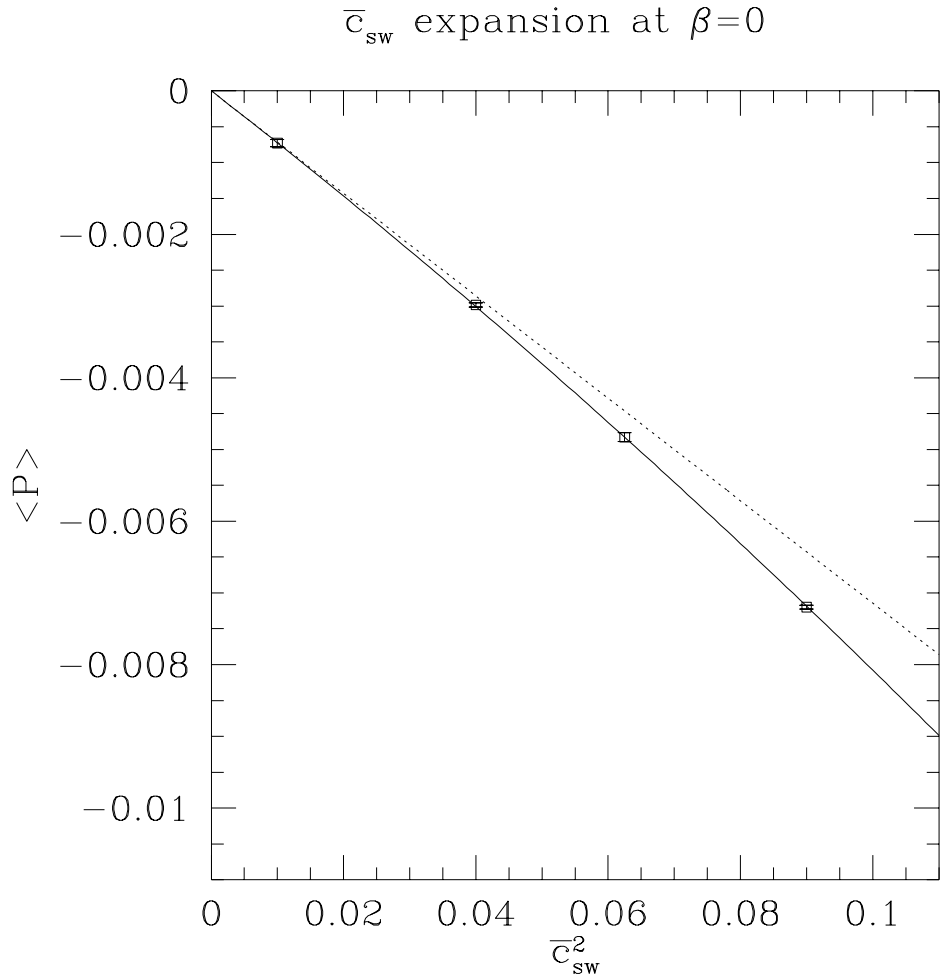}}
\vspace{0.0cm}
\begin{center}
\parbox{13cm}{\caption{ \label{fig:cumulant_0}
 \figfont
 We plot the average plaquette value 
 as measured from
 our Monte Carlo simulations for various values of
 $\bar{c}^2_{sw}$ at $\beta=0$ on a $4^4$ lattice. 
 This is to be compared with the analytic result obtained
 from the small $\bar{c}_{sw}$ expansion which is plotted
 as the lines. The dotted line indicates the result of the
 leading order (eq.~(\ref{eq:p2})) while the solid line also includes the next 
 leading order in the expansion (eq.~(\ref{eq:p4})). 
}}
\vspace{1.0cm}
\end{center}
\end{figure}
some more work. Basically one is led to the following general expression:
\ba
\label{eq:plq}
<P>&=& 
\bar{c}^2_{sw} <P Tr(\sigma {\cal F})^2>_0
-{\bar{c}^4_{sw} \over 2}<P Tr(\sigma{\cal F})^4>_0
\nonumber \\
&+& {\bar{c}^4_{sw} \over 2}<P Tr(\sigma{\cal F})^2Tr(\sigma{\cal F})^2>_0
\nonumber \\
 &-&\bar{c}^4_{sw}<P Tr(\sigma{\cal F})^2>_0<Tr(\sigma{\cal F})^2>_0\;\;,
\ea
with $<O>_0=\int dU O(U)$ and $dU$ being the normalized group measure.
In the above formula, we have used the shorthanded notation
$\sigma {\cal F}$ to stand for $\sigma_{\mu\nu}{\cal F}_{\mu\nu}/2$.
The last two terms correspond to the connected contribution of
the operators $P$ and $Tr(\sigma{\cal F})^2$.
The calculation of these terms is quite straightforward but rather
technical. We list some of the details in the appendix.
This calculation results in: 
\be
\label{eq:p4}
<P> = 
- {1 \over 12}\bar{c}_{sw}^2(1 + 
{475 \over 384}\bar{c}^2_{sw})\;\;.
\ee

We made simulations with SW-improved action 
in the Schr\"odinger functional setup \footnote{\footfont For the
timelike plaquettes touching the boundary within the 
Schr\"odinger functional, eq.(\ref{eq:p4}) is replaced by 
$- (1 / 24)\bar{c}_{sw}^2(1 + (491 / 384)\bar{c}^2_{sw})$. }
at various values of $\bar{c}_{sw}$ at
$\beta=0$ on a $4^4$ lattice and the
results 
 are shown in Fig.~\ref{fig:cumulant_0} together with
the prediction of eq.~(\ref{eq:p2}) and eq.~(\ref{eq:p4}).
For small enough values of $\bar{c}^2_{sw}$ (less than $0.01$), 
the lowest order
of the 
expansion is already satisfactory. For larger $\bar{c}^2_{sw}$
values, the deviation of the data from the lowest order prediction is quite 
significant. However, if one also includes the second order contribution, 
all 
data points are in good agreement with the anticipated values 
up to $\bar{c}^2_{sw} \sim 0.1$.

\subsection{Tests at nonvanishing $\beta$}

For a nonvanishing value
of $\beta$
and any observable $O$, one
can obtain an expansion in terms of numerically
computed cumulants at
$\bar{c}_{sw}=0$. We have
\ba
\label{eq:cumulant}
\!\!\!<O>&=& <O>_0
+2\bar{c}_{sw}^2 \left[ <O Tr({\cal F}_{\mu\nu}{\cal F}_{\mu\nu})>_0
-<O>_0<Tr({\cal F}_{\mu\nu}{\cal F}_{\mu\nu})>_0 \right]
\nonumber \\
  &+&\!\!\!{8 \over 3}\bar{c}_{sw}^3 \left[
<O Tr({\cal F}_{\mu\nu}{\cal F}_{\nu\beta}{\cal F}_{\beta\mu})>_0
  -<O>_0 <Tr({\cal F}_{\mu\nu}{\cal F}_{\nu\beta}
{\cal F}_{\beta\mu})>_0 \right].
\ea
\begin{figure}[htp]
\vspace{-0mm}
\centerline{ \epsfysize=13.5cm
             \epsfxsize=13.5cm
             \epsfbox{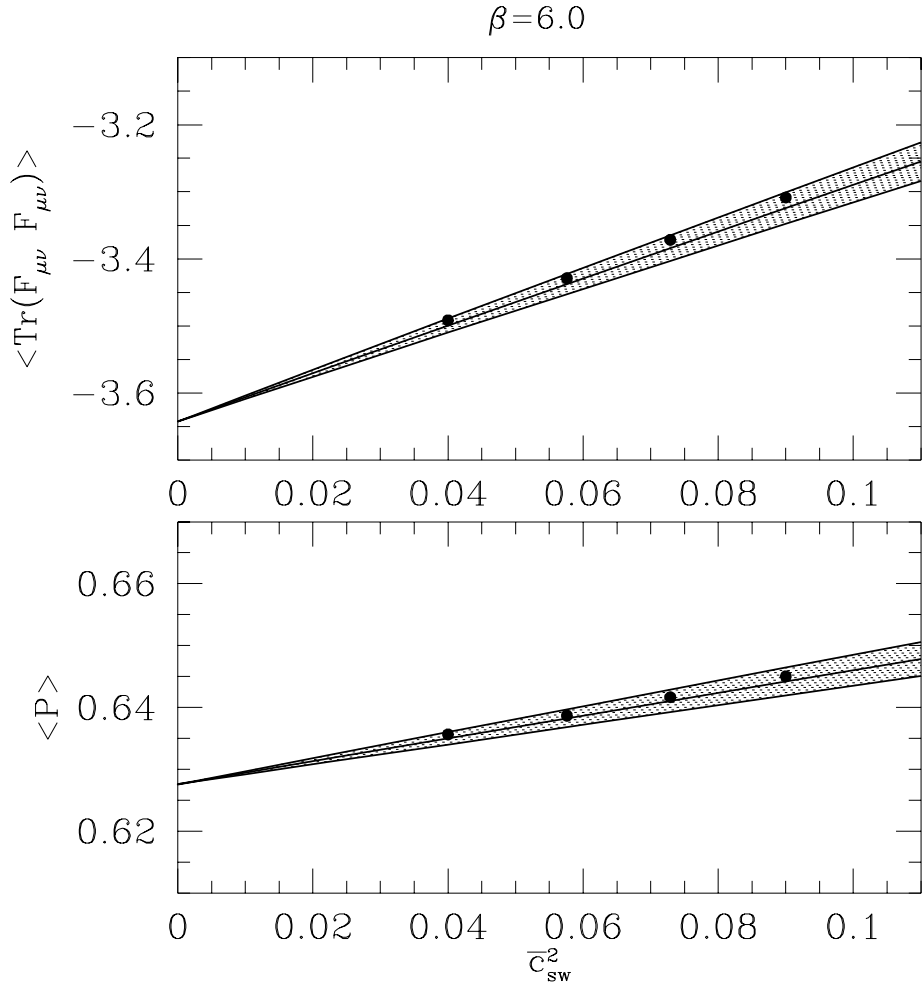}}
\vspace{-0mm}
\begin{center}
\parbox{13cm}{\caption{ \label{fig:cumulant}
 \figfont
 We plot the average plaquette value and the quantity
 $Tr({\cal F}_{\mu\nu}{\cal F}_{\mu\nu})$ as measured from
 our Monte Carlo simulations for various values of
 $\bar{c}^2_{sw}$. All of these data points stay within
 the shaded region as predicted by the cumulant expansion,
 which is constructed from a high statistics run ($5.5$ million
 measurements) at
 $\bar{c}_{sw}=0$. The shaded bands represent the
 error of the cumulant expansion obtained by analyzing
 the blocked errors of the various cumulants at
 $\bar{c}_{sw}=0$.
}}
\end{center}
\end{figure}
In the above formula, the notation $<...>_0$ stands now for
the Monte Carlo average at $\bar{c}_{sw}=0$.
We have performed a high statistics run at a fixed $\beta$ value
with $\bar{c}_{sw}=0$, measured the cumulants above and constructed the 
expansion. On the other hand, we have made simulations at a
nonvanishing value of $\bar{c}_{sw}$ directly 
and compared with the cumulant expansion.
This test was performed at $\beta=6$ on
a $4^4$ lattice and the result is illustrated in
Fig.~\ref{fig:cumulant}. 

Obviously, for the two quantities that have
been measured, namely the average plaquette value and the average
value of $Tr({\cal F}_{\mu\nu}{\cal F}_{\mu\nu})$, the measured values from the
simulation agree very well with the prediction of the cumulant 
expansion. The width of the shaded bands in the figure represents 
the statistical uncertainty of the cumulant expansion which
is constructed from a run at $\bar{c}_{sw}=0$ with
$5.5$ million measurements. In this analysis, we only
took into account the blocked  statistical errors of the run at
$\bar{c}_{sw}=0$. The $O(\bar{c}^3_{sw})$ term in the cumulant
expansion~(\ref{eq:cumulant}) is also included, the effect
of which is however rather small. It is clear from the figure that
the systematic errors in this case should be smaller 
compared with the statistical errors 
for the parameter range we are considering here.
Of course, by measuring more cumulants
at $\bar{c}_{sw} = 0$, 
one could also construct higher order corrections in the cumulant
expansion. 

\section{Timing of the Program}

When comparing with the conventional Wilson QCD simulation, 
the timing of the version with
the improved fermion action is a crucial piece of information.
\begin{figure}[htp]
\vspace{-0mm}
\centerline{ \epsfysize=13.5cm
             \epsfxsize=13.5cm
             \epsfbox{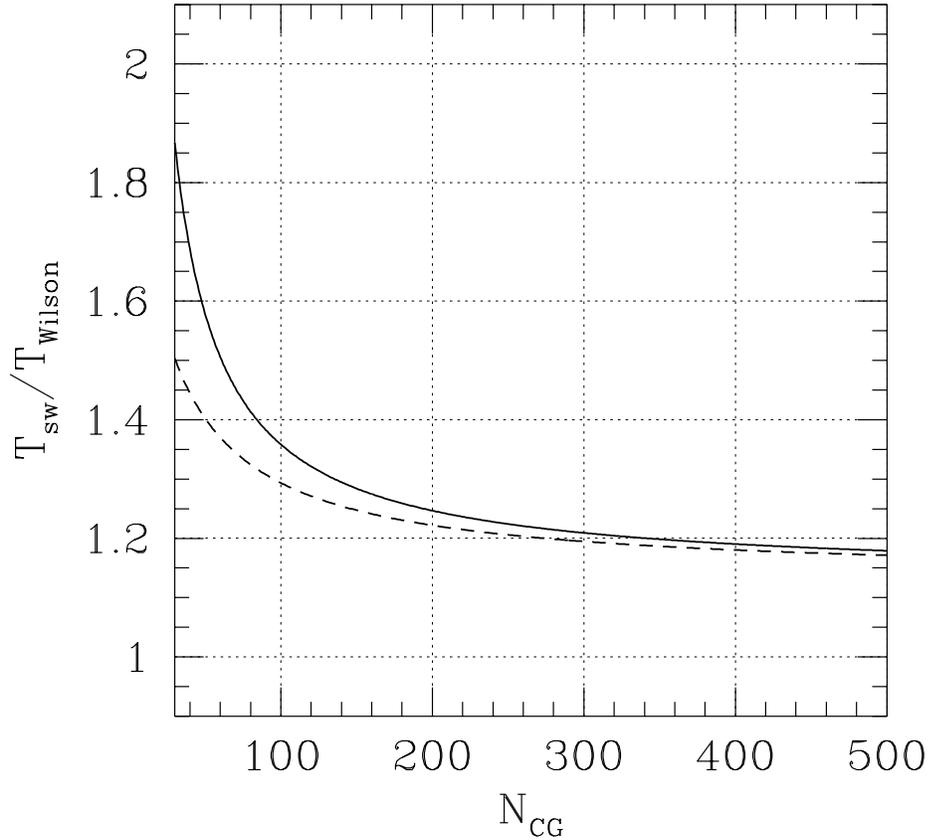}}
\vspace{-0mm}
\begin{center}
\parbox{13cm}{\caption{ \label{fig:time}
 \figfont
 The CPU time for the algorithm of the improved action 
 is plotted as a function of the number of conjugate
 gradient iterations per step ($N_{CG}$) relative to
 the conventional Wilson fermion simulation. 
 The solid and the dashed lines correspond to the choice 
 of $c^{(0)}_w=0$ and $c^{(0)}_w=50$ in eq.~(\ref{eq:time}) 
 respectively.
 We see that, in the regime of $N_{CG}$ values expected for QCD simulations,
 the algorithm for the improved action
 takes about 20 percent more time
 to generate a configuration.
}}
\end{center}
\end{figure}
In a typical QCD simulation using molecular dynamics
algorithms, 
 the conjugate gradient (CG) iterations dominate 
the CPU time of the program. 
Therefore, in the limit of number of conjugate
gradient iterations $N_{CG}$ going to infinity, 
 the time of the matrix-vector
multiplication is the most crucial part. 
In our version of Symanzik improvement, 
this multiplication turns out to be
only slightly slower than the conventional
Wilson case by about 15 percent. 
The force evaluation, in this case, adds some
overhead to the program which does not
depend on $N_{CG}$. It is 
convenient to express all times in units of the
matrix-vector multiplication of the
conventional Wilson case. 

To be specific, we propose the 
following formula for the time of the program:
\ba
\label{eq:time}
T_{Wilson} &=& c^{(0)}_w+(c^{(1)}_w+c^{(2)}_w)N_{CG} \;\;,
\nonumber \\
T_{SW} &=& c^{(0)}_{sw}+(c^{(1)}_{sw}+c^{(2)}_{sw})N_{CG} \;\;.
\ea
In the above formula, the coefficient $c^{(2)}_w=2$ is, by definition, the
number of matrix-vector 
multiplications needed for each CG iteration in
the conventional Wilson case. The coefficient $c^{(1)}_w$ represents
the cost of other operations for each CG iteration (linear combinations, 
inner products etc.). Typically, this coefficient is quite
small. The coefficient $c^{(0)}_w$ is  the 
overhead that does not depend on  $N_{CG}$. The value of
$c^{(0)}_w$ depends on the implementation of the program.
However, in the asymptotic region where $N_{CG}$ is large,
the effect of $c^{(0)}_w$ becomes irrelevant.

The quantities $c^{(0)}_{sw}$, $c^{(1)}_{sw}$ and $c^{(2)}_{sw}$  
are the corresponding coefficients for the program with 
the Sheikholeslami-Wohlert action.
In this case, $c^{(2)}_{sw}=2.3$ is slightly larger than 
the corresponding value in the Wilson case. The value of
$c^{(1)}_{sw}=c^{(1)}_{w}$ remains the same and 
we find, for our implementation of the SW-term on the APE computer, 
the coefficient 
$c^{(0)}_{sw}\sim (50+c^{(0)}_w)$ to be larger than that of
the Wilson case. 
In Fig.~\ref{fig:time} we plot the CPU time of our program
with improved fermions relative to the
conventional Wilson case as a function
of the  number of CG-iterations per molecular dynamics step.  
The solid and dashed curves in the figure correspond to
the choice of $c^{(0)}_w=0$ and $c^{(0)}_w=50$ respectively, which
we consider to be two rather extreme cases.
For typical simulations, the curve should lie between these two.
In any case, we see that, for the most
interesting physical situation in which  
$N_{CG} \sim 200$ or more, the simulation with improved fermions
is only about 20\%  slower, quite independent
of the value of $c^{(0)}_w$. 

\section{Conclusions}

We demonstrated in this paper that the SW-term as required by Symanzik's
improvement program to cancel $O(a)$ effects in Wilson QCD  
can be implemented straightforwardly for simulation
algorithms like Hybrid Monte Carlo or the Kramers equation. For the 
implementation even-odd preconditioning can
be maintained which is an important aspect concerning the performance of 
the algorithms. 
We designed tests for checking the implementation
of the SW-term by analytical and numerical computations
of cumulant expansions. 

The most important outcome of our work is illustrated in Fig.~6
which compares the time of a program with conventional
Wilson fermions with the one where the SW-term has been
added. The figure nicely demonstrates that the overhead due to the
SW-term in the lattice action is only about 20$\%$ when the number
of conjugate gradient iterations exceeds about 200 iterations, which
is a number easily reached 
in realistic simulations of lattice QCD.
On the other hand, 
Symanzik improvement will enable us to perform 
QCD simulations with much smaller lattice artifacts than 
the conventional Wilson fermions.
Since the cost of a QCD simulation is proportional to
a large power of 
 $a^{-1}$,
 any small factor of improvement in lattice spacing will reduce the cost of
the simulation significantly. 

This brings us to the conclusion that simulations with SW-improved
dynamical Wilson fermions can be done without much more cost
than conventional Wilson fermion simulations. We expect that
the cancellation of 
the strong $O(a)$ effects, as 
have been seen in quenched simulations,
will by far merit 
the small overhead due to the addition of the SW-term.

A last remark concerns the multiboson technique for simulating lattice
QCD \cite{luscher,multiboese}. This relatively new method shows a comparable
performance to the molecular dynamics algorithms 
and can be considered as a real 
alternative to the Hybrid Monte Carlo or Kramers equation algorithms.
However, for the multiboson technique
it is not straightforward to implement the SW-term, if one wants
to use heatbath or over-relaxation algorithms. 
We propose therefore, 
to use exact local versions  
of the Hybrid Monte Carlo or
Kramers equation algorithms \cite{localhmc} 
for the simulation. For these algorithms,
the implementation of the SW-term described here directly
applies. 

\section{Acknowledgements}

We would like to thank 
M. L\"uscher and U. Wolff for very helpful discussions.
We thank P. Weisz and U. Wolff for offering their
matrix multiplication routines for the preconditioned fermion matrix
to us. The numerical simulations have been performed on the 
APE/Quadrics computers at DESY-IFH (Zeuthen). We thank the staff of the
computer center at Zeuthen for their support.

\newpage
\appendix
\section{Analytic computation of the cumulant expansion at $\beta=0$ }

In this appendix, we discuss the analytic calculation of the
cumulant expansion for the average plaquette value at
$\beta=0$ in some detail.
Our aim is to calculate the ensemble average of 
the plaquette which is defined as
$P=(1/6)tr(U_P+U^{\dagger}_P)$.
We will only consider the case 
of periodic boundary conditions for the gauge fields here.
The plaquette is specified by its (geometrical) ``orientation'', which
designates the plane that the plaquette lies in, and its two possible 
``circulations'', 
one corresponding to $U_P$ and the other to $U^{\dagger}_P$.
When we expand the effective action~(\ref{eq:seffb0}) into 
power series of $\bar{c}_{sw}$, at each order,
we will have products of plaquettes of various
orientations and circulations.
However, in order to obtain a nonvanishing contribution after
group integrations,
all the plaquettes that appear in
the product have to be ``tiled'' according to
the following simple rule: 
\begin{list}{  }
\item{For any plaquette appearing in the
expansion, the number of factors of one circulation minus
the number of factors of the opposite circulation
has to be $0$ (mod 3).}
\end{list}


\noindent The average plaquette value, up to $O(\bar{c}^4_{sw})$, is
given by equation~(\ref{eq:plq}) in the main text.
To obtain the contribution of $O(\bar{c}^4_{sw})$, one has
to evaluate the contributions from the following two cases:
(1) the connected contributions of $P$,   
$Tr(\sigma {\cal F})^2$ and $Tr(\sigma {\cal F})^2$ and
(2) the contributions from  $PTr(\sigma {\cal F})^4$.
We will now study these two cases separately.
Since $tr(U_P)$ and $tr(U^{\dagger}_P)$  always
give the same contribution, we will just consider
the former. In the calculation below, the following
common factor always appears:
\be
c_{fac} = 2 \cdot {1 \over 6} \cdot {\bar{c}^2_{sw} \over 64} 
\cdot {\bar{c}^2_{sw} \over 64} \;\;,
\ee
which comes from the definitions of the plaquette and
${\cal F}_{\mu\nu}$.

\subsection{The connected contribution  
$<PTr(\sigma {\cal F})^2Tr(\sigma {\cal F})^2>_{c0}$}
 
The connected contribution 
$<PTr(\sigma {\cal F})^2Tr(\sigma {\cal F})^2>_{c0}$ is
defined to be
\ba
<PTr(\sigma {\cal F})^2Tr(\sigma {\cal F})^2>_{c0}
&=&<PTr(\sigma {\cal F})^2Tr(\sigma {\cal F})^2>_{0}
\nonumber \\
&-&2<PTr(\sigma {\cal F})^2>_0<Tr(\sigma {\cal F})^2>_{0} \;\;.
\ea
On the right hand side of the above equation, the first
term receives contributions of the type
$tr(U_P)tr(U_PU_P)tr(U_{P'}U^{\dagger}_{P'})$ which
exactly cancels the second term. The remaining contributions 
 only come in from two different
cases which we will call (a) and (b). In  case (a), 
the two factors of $Tr(\sigma {\cal F})^2$ are sitting
at neighboring corners of the plaquette to be measured.
The link which connects these two corners is also
an edge of another plaquette $P' \neq P$ which has the same
geometrical orientation.
Then, the combination of the type 
$tr(U_P)tr(U_PU_{P'})tr(U_PU^{\dagger}_{P'})$ 
fulfills the rule and gives a contribution of
\be
<PTr(\sigma {\cal F})^2Tr(\sigma {\cal F})^2>^{(a)}_{c0} =
{512\over 3} \cdot c_{fac}.
\ee

In case (b), the situation is the same as in
case (a) except that two factors of $Tr(\sigma {\cal F})^2$ 
are sitting at the same corner of
the plaquette to be measured. Now, since we have
$3$ different ways of choosing  the other plaquette $P'$, 
we get a contribution:
\be
<PTr(\sigma {\cal F})^2Tr(\sigma {\cal F})^2>^{(b)}_{c0} =
3 <PTr(\sigma {\cal F})^2Tr(\sigma {\cal F})^2>^{(a)}_{c0} 
\;\;.
\ee
This concludes the calculation of the contribution
 $<PTr(\sigma {\cal F})^2Tr(\sigma {\cal F})^2>_{c0}$.

\subsection{The contribution from $<PTr(\sigma {\cal F})^4>_0$}

In this case, there could be two possibilities concerning the
Dirac structure of the factors from $\sigma {\cal F}$: 
\begin{list}{  }
 \item {(A) All $4$ factors have the same geometrical orientation as the
  plaquette to be measured.}
 \item {(B) Two of the $4$ factors have the same geometrical orientation as the
  plaquette to be measured, the other two take orthogonal orientations.}
 \vspace{3mm}
\end{list}
\noindent We will now discuss these contributions separately.

For case (A), we have the following $3$ possibilities which we
will list as (A1), (A2) and (A3). In all these cases, the 
Dirac trace simply gives a factor of $4$.
In case (A1), we consider the combination $tr(U_PU_PU_{P'}U^{\dagger}_{P'})$
, $P\ne P'$.
Note that
there are $4$ such terms in $Tr(\sigma{\cal F})^4$. The contribution is:
\be
<PTr(\sigma {\cal F})^4>^{(A1)}_0 =
384 \cdot c_{fac}\;\;.
\ee
 
In case (A2), the situation is almost the same as in case (A1) with
the exception that the structure is 
of the type $tr(U_PU_{P'}U_PU^{\dagger}_{P'})$.
Since there are $2$ such terms in $Tr(\sigma{\cal F})^4$, 
the contribution is:
\be
<PTr(\sigma {\cal F})^4>^{(A2)}_0 =
-64 \cdot c_{fac}\;\;.
\ee

In case (A3), at least one of the $\sigma {\cal F}$ factors contributes 
a $U^{\dagger}_P$ which has a circulation opposite to
that of the plaquette to be measured.
This contribution is given by:
\be
<PTr(\sigma {\cal F})^4>^{(A3)}_0 =
16 \cdot c_{fac} \cdot I\;\;,
\ee
where the group integral $I$ is given by
\be \label{i-integral}
I = \int dU tr(U) tr[(U-U^{\dagger})^4]= \int dU  tr(U)tr[(U^{\dagger})^4-4U^2]
\;\;.
\ee
This integral, as well as the other group integrals appearing
in the different contributions, can 
be evaluated using the techniques described
in appendix B. The value of $I$ in (\ref{i-integral}) equals $5$.
This concludes the discussion of case (A).

For the case (B), we have the following two possibilities which
we call (B1) and (B2). 

In case (B1), the situation is very similar
to case (A1) above, i.e. the contribution has the form
$tr(U_PU_PU_{P'}U^{\dagger}_{P'})$. 
 In this case, the Dirac trace will just give a factor of $4$. 
The two factors  $U_{P'}$ and $U^{\dagger}_{P'}$ can 
take the $5$ orientations
orthogonal to the one specified by the measured plaquette.
They can also take all $4$ possible leaves and $2$ possible
circulations in the clover.
This contribution turns out to be the dominant
one at this order. We have:
\be
<PTr(\sigma {\cal F})^4>^{(B1)}_0 =
2560 \cdot c_{fac}\;\;.
\ee

Case (B2) is closely related to (A2) and the structure is
$tr(U_PU_{P'}U_PU^{\dagger}_{P'})$.
However, since $P'$ does not have the same geometrical orientation 
as $P$, 
the Dirac trace should be done with more care. Basically, for
any given orientation of the measured plaquette, 
among the other $5$ orthogonal orientations,
only one of them gives a Dirac trace of $4$ while the other $4$ orientations
give a Dirac trace of $(-4)$ due to the anticommuting properties of
the $\sigma$ matrices. Therefore, effectively there appear to
be only $3$ orientations to contribute a Dirac trace 
 $(-4)$. We get 
\be
<PTr(\sigma {\cal F})^4>^{(B2)}_0 =
256 \cdot c_{fac}\;\;.
\ee
This concludes our calculation for case (B).

Adding all contributions up, gives the result quoted in the main text.

\section{Evaluation of group integrals for $SU(3)$}

Here we discuss some techniques to evaluate the group integrals
for the group $SU(3)$. The integrals that we wish to evaluate 
are of the form
\be
\int dU [tr(U)]^{i_1}[tr(U^{-1})]^{j_1}[tr(U^2)]^{i_2}[tr(U^{-2})]^{j_2}\cdots
\;\;,
\ee
where $U$ is an element of $SU(3)$ and the measure $dU$ stands for
the normalized Haar measure of the group.
The first step is to use the Cayley relation for an $SU(3)$ matrix:
\be
U^3 = tr(U)U^2 - tr(U^{-1})U + 1\;\;,
\ee
which could be established easily from the fact that
$(U-\lambda_1)(U-\lambda_2)(U-\lambda_3)=0$, where
$\lambda_1$, $\lambda_2$ and $\lambda_3$ are the
eigenvalues of $U$ and unit matrices are to be substituted appropriately.
With the help of this relation, we can transform 
terms with powers of $U$ higher than $1$ into powers
of $1$, $0$ and $(-1)$. This means that, without loss
of generality, we can discuss the following integral:
\be
\int dU [tr(U)]^{i_1}[tr(U^{-1})]^{j_1}
\;\;.
\ee
This integral can be done by realizing that the integrand above is
nothing but the character of the representation 
$\mathbf{3}^{i_1} \otimes \bar{\mathbf{3}}^{j_1}$, 
assuming $U$ is in the $\mathbf{3}$ representation.
Therefore, the above integral is equal to the number of trivial
representations in this direct product. All one has to do now is
to decompose the direct product 
$\mathbf{3}^{i_1} \otimes \bar{\mathbf{3}}^{j_1}$ 
and count how many times the 
trivial representation appears.

As an example, let us consider the integral
\be
\int dU [tr(U^{-1})][tr(U^4)]
\;\;.
\ee
Using Cayley relation, this is transformed into: 
\be
\int dU \left[ tr(U^{-1})tr(U)^4-4 tr(U^{-1})^2tr(U)^2
+2 tr(U^{-1})^3+4 tr(U^{-1})tr(U) \right]
\;\;.
\ee
The last term just gives $4$ and the third term gives $2$.
One easily finds out that $\mathbf{3}\otimes \mathbf{3} 
\otimes\bar{\mathbf{3}}\otimes \bar{\mathbf{3}}$
contains $2$ trivial representations. So the second term yields
$(-4)\cdot 2=(-8)$. The first term requires the decomposition of
$\bar{\mathbf{3}}\otimes \mathbf{3}
\otimes \mathbf{3}\otimes \mathbf{3}\otimes \mathbf{3}$ and 
one finds that it contributes a $3$.     
Therefore the integral finally yields a value of $1$.


\vspace{-1mm}
\begin{thebibliography}{9}
\settowidth{\baselineskip}{\reffont QC}
%
\bibitem{letter} K.~Jansen, C.~Liu, M.~L\"{u}scher, H.~Simma, S.~Sint,
R.~Sommer, P.~Weisz and U.~Wolff, 
DESY-preprint, DESY 95-230, hep-lat/9512009.
%
\bibitem{symanzik} K.~Symanzik, Some Topics in quantum field theory, in:
Mathematical problems in theoretical physics, eds. R. Schrader et. al.,
Lecture Notes in Physics, Vol. 153 (Springer, New York, 1982);
Nucl.Phys.B226 (1987) 187 and 205.
%
\bibitem{clover} B. Sheikholeslami and R. Wohlert, Nucl. Phys. B259
 (1985) 572.
%
\bibitem{quench} 
D.K. Sinclair, 
Plenary review talk presented at LATTICE 1995,  Melbourne, 11th - 15th
July, 1995, 
Argonne preprint, ANL-HEP-CP-95-63, to appear in the Proceedings.
%
\bibitem{sf} M. L\"uscher, R. Narayanan, P. Weisz and U. Wolff, 
Nucl.Phys.B 384 (1992) 168; \\
S. Sint, Nucl.Phys.B421 (1994) 135, Nucl.Phys.B451 (1995) 416.
%
\bibitem{tony} S. Duane, A.D. Kennedy, B.J. Pendleton and D. Roweth,
 Phys. Lett. B195 (1987) 216.
%
\bibitem{horowitz}
A.~M.~Horowitz, Phys. Lett. 156B (1985) 89; Nucl. Phys. B280 (1987) 510;
               Phys. Lett. 268B (1991) 247.
%
\bibitem{kramers} K. Jansen and C. Liu, Nucl. Phys. B453 (1995) 375.
%
\bibitem{precond}
T.~Degrand and P.~Rossi, Comp. Phys. Comm. 60  (1990) 211.
%
\bibitem{ssor} S. Fischer, A. Frommer, U. Gl\"assner, T. Lippert, 
G. Ritzenh\"ofer and K. Schilling, 
HLRZ preprint, HLRZ 4/96, hep-lat/9602019.
%
%
\bibitem{luo} X.-Q. Luo, Comp. Phys. Comm. 94 (1996) 119.
%
\bibitem{house} D.~W.~Stewart, Introduction to Matrix Computations,
Academic Press, Inc., 1973.
%
\bibitem{ulli} P. Weisz and U. Wolff, DESY internal report, unpublished.
%
\bibitem{allton}
C. R. Allton et al., UKQCD collaboration, Phys. Rev.
D 49 (1994) 474.
%
\bibitem{sugar} S. Gottlieb, W. Liu, D. Toussaint, R. L. Renken and
R. L. Sugar, Phys. Rev. D 35 (1987) 2531.
%
\bibitem{luscher}
M.~L\"uscher, Nucl. Phys. B418 (1994) 637.
%
\bibitem{multiboese}
B.~Bunk, K.~Jansen, B.~Jegerlehner, M.~L\"uscher,
H.~Simma and R.~Sommer, hep-lat/9411016;
Nucl. Phys. B (Proc. Suppl.) 42 (1995) 49;\\
B.~Jegerlehner, hep-lat/9411065;
Nucl. Phys. B (Proc. Suppl.) 42 (1995) 879; \\
M. Peardon, Nucl. Phys. B (Proc. Suppl.) 42 (1995) 891, hep-lat/9412008;\\
%
A. Borici and P. de Forcrand, ETH Z\"urich preprint, IPS-95-13,
hep-lat/950502;\\
%
C. Alexandrou, A. Borrelli, P. de Forcrand, A. Galli and F. Jegerlehner,
PSI preprint, PSI-PR-95-09, hep-lat/9506001;\\
%
P. de Forcrand, ETH Z\"urich preprint, IPS-95-24,
hep-lat/9509082;\\
%
B. Jegerlehner, MPI preprint, MPI-PhT 95-119,
hep-lat/9512001;\\
%
K. Jansen, B.  Jegerlehner and C. Liu, DESY preprint, 
DESY 95-243, hep-lat/9512018;\\
%
A. Borrelli, P. de Forcrand, A. Galli
MPI Munich preprint, hep-lat/9602016.               
%
\bibitem{localhmc} A.D. Kennedy and K.M. Bitar, Nucl.Phys. (Proc.Suppl.)
34 (1994) 786;\\
P. Marenzoni, L. Pugnetti and P. Rossi, Phys.Lett. B315 (1993) 152.
%

\end{thebibliography}
\end{document}